\documentclass{aastex}

\RequirePackage{color}

\begin{document}

\title{Compact models with regular charge distributions}
\slugcomment{}
%% Running heads
\shorttitle{Short article title}
\shortauthors{Autors et al.}

\author{P. Mafa Takisa\altaffilmark{1}} \and \author{S. D. Maharaj\altaffilmark{1}}
\affil{Astrophysics and Cosmology Research Unit,
 School of Mathematics, Statistics and Computer Science,
 University of KwaZulu-Natal, Private Bag X54001,
 Durban 4000, South Africa}

%\email{\emaila}

\begin{abstract}
We model a compact relativistic body with anisotropic pressures in the presence of an electric field. The 
equation of state is barotropic with a linear relationship between the radial pressure and the 
energy density. Simple exact models of the 
Einstein-Maxwell equations are generated. A graphical analysis 
indicates that the matter and electromagnetic variables are well behaved. In particular the proper 
charge density is regular for certain parameter values at the 
stellar centre unlike earlier anisotropic models in the presence of 
charge. We show that the electric field affects the mass of stellar objects and the observed mass for a 
particular binary pulsar is regained. Our models contain 
previous results of anisotropic charged matter with a linear equation 
of state for special parameter values.
\end{abstract}


%\section*{}
%\label{sec:intro}

\section{Introduction \label{intro}}
Solutions of the Einstein-Maxwell system of equations for static spherically symmetric interior 
spacetimes are important in describing charged compact objects 
in relativistic astrophysics where the gravitational field is strong, as 
in the case of neutron stars. The detailed analyses of \cite{1} and \cite{2}
show that the presence of electromagnetic field affects the 
values of redships, luminosities and maximum mass of compact 
objects. The role of the electromagnetic field in describing the gravitational behaviour of stars composed 
of quark matter has been recently highlighted by \cite{3} and \cite{4,5}. In recent 
years, many researchers have attempted to introduce different approaches of finding solutions to the 
field equations. \cite{6} found 
solutions to the Einstein-Maxwell system with a specified form of the 
electrical field with isotropic pressures. These solutions satisfy a barotropic equation of state and regain 
the \cite{7} model. \cite{8} found new exact classes of solutions to the Einstein-Maxwell system. They 
considered anisotropic pressures in the presence of the electromagnetic field 
with the linear equation of state of strange stars with 
quark matter. Other recent investigations involving charged relativistic 
stars include the results of \cite{9}, \cite{10} and \cite{11}. 
The approach of \cite{11a} is interesting in that it utilizes the existence of 
a conformal symmetry in the spacetime manifold to find a solution. These exact solutions are 
relevant in the description of dense relativistic astrophysical objects. 
Applications of charged relativistic models include studies in cold compact objects by \cite{12}, analyses of strange matter and binary 
pulsars by \cite{13}, and analyses of quark-diquark mixtures in equilibrum by \cite{14}. \cite{15},
\cite{16} and \cite{17}
modeled charged core-envelope stellar structures in which the core 
of the sphere is an isotropic fluid surrounded by a layer of anisotropic fluid. To get more flexibility in 
solving the Einstein-Maxwell system, \cite{18} considered a general approach of dealing with anisotropic 
charged matter with linear or nonlinear equations of state. Their approach offers a fresh view of 
relationships between the equation of state, charge 
distributions and pressure anisotropy. A detailed physical analysis showed the 
desirability of models with an equation of state. However there are few models in the literature with 
an equation of state which satisfy the physical criteria.

It is desirable that the criteria for physical acceptability, as given in \cite{19}, be satisfied in a realistic charged stellar 
model. In particular the proper charge density should 
be regular at the centre of the sphere. This is an essential 
requirement for a well behaved electromagnetic field, and this important feature has been highlighted in 
the treatment of \cite{18}. In many 
models found in the past, including the charged anisotropic solution 
with a linear equation of state of \cite{8}, the proper charge density is 
singular at the centre of the star. For a realistic description this is a limiting feature for the applicability
of the model. It is desirable to avoid this 
singularity if possible. Our object in this paper is to generate new  
solutions to the Einstein-Maxwell system which satisfy the physical properties: the gravitational 
potentials, electric field intensity, charge distribution and matter 
distribution should be well behaved and regular throughout the star, 
in particular at the stellar centre.
We find new exact solutions for  a charged relativistic sphere with anisotropic pressures and a linear 
equation of state. Previous models are shown to be 
special cases of our general result.
In section \ref{Basic}, we rewrite the Einstein-Maxwell field equation for a static spherically symmetric 
line element as an equivalent set of differential equations 
using a transformation due to \cite{20}. In section  \ref{choice}, 
we motivate the choice of the gravitational potential and the electric field intensity that enables us to integrate the field equations.
In section \ref{models}, we present new exact solutions to the Einstein-Maxwell system, which contain 
earlier results. The singularity in the charge density may 
be avoided.
In section \ref{features}, a physical analysis of the new solutions is performed; the matter variables and 
the electromagnetic quantities are plotted. Values for 
the stellar mass are generated for charged and uncharged matter for 
particular parameter values. The values are consistent with the conclusions of \cite{21,22,23}
for strange stars. The variation of density for different charged compact structures is discussed in section \ref{Q}.
We summarise the results obtained in this paper in section \ref{concl}.

\section{Basic equations\label{Basic}}
In standard coordinates the line element for a static spherically symmetric fluid has the form
\begin{equation}
\label{eq:f1} {\rm d} s^{2} = -e^{2\nu(r)}  {\rm d} t^{2} + e^{2\lambda(r)} {\rm d} r^{2} + r^{2}( {\rm d} 
\theta^{2} + \sin^{2}{\theta} {\rm d} \phi^{2}).
\end{equation}
The Einstein-Maxwell equations take the form
\begin{eqnarray}
\label{eq:f2}
\frac{1}{r^{2}}\big[r(1-e^{-2\lambda})\big]^{\prime} = \rho+\frac{1}{2}E^{2},\\
\label{P24b}
-\frac{1}{r^{2}}(1-e^{-2\lambda})+\frac{2\nu^{\prime}}{r}e^{-2\lambda} = p_{r}-\frac{1}{2}E^{2},\\
\label{P24c}
e^{-2\lambda}\bigg(\nu^{\prime\prime}+\nu^{\prime2}+\frac{\nu^{\prime}}{r}-\nu^{\prime}\lambda^
{\prime}-\frac{\lambda^{\prime}}{r}\bigg) = p_{t}+\frac{1}
{2}E^{2},\\
\label{P24d}
\sigma = \frac{1}{r^{2}}e^{-\lambda}(r^{2}E)^{\prime},
\end{eqnarray}
where dots represent differentiation with respect to \textit{r}.
It is convenient to introduce a new 
independent variable \textit{x} and introduce new functions 
\textit{y} and \textit{Z}:
\begin{equation}
\label{eq:f3} x = Cr^2,~~ Z(x)  = e^{-2\lambda(r)}, ~~
A^{2}y^{2}(x) = e^{2\nu(r)},
\end{equation}
where \textit{A} and \textit{C} are constants. We assume a barotropic equation of state
\begin{equation}
\label{eq:f4} p_r = \alpha \rho - \beta,
\end{equation}
relating the radial pressure $p_{r}$ to the energy density $\rho$. The quantity $p_{t}$ is the tangential 
pressure, \textit{E} represents the electric field intensity, 
and $\sigma$ is the proper charge density. Then the equations 
governing the gravitational behaviour of a charge anisotropic sphere, with a linear equation of state, are 
given by
\begin{eqnarray}
\label{eq:f5}
 \frac{\rho}{C} = \frac{1-Z}{x}-2\dot{Z}-\frac{E^{2}}{2C},\\
\label{eq:f6}
 p_{r}  =  \alpha\rho-\beta,\\
\label{eq:f7}
p_{t}  =  p_{r}+\Delta,\\
\label{eq:f8}
\Delta  =  4CxZ \frac{\ddot{y}}{y}+2C\left[x\dot{Z}+\frac{4Z}{(1+\alpha)}\right]\frac{\dot{y}}{y} \nonumber\\
+\frac
{(1+5\alpha)}{(1+\alpha)}C\dot{Z} 
 -\frac{C(1-Z)}{x}+\frac{2\beta}{(1+\alpha)},\\
\label{eq:f9}
 \frac{E^{2}}{2C}  =  \frac{1-Z}{x}-\frac{1}{(1+\alpha)}\left[2\alpha\dot{Z}+4Z\frac{\dot{y}}{y}+\frac
{\beta}{C} \right],\\
\label{eq:f10}
\frac{\sigma^{2}}{C}  = \frac{4Z}{x}\left(x\dot{E}+E\right)^{2},
\end{eqnarray}
where $\Delta = p_t -p_r$ is called the measure of anisotropy. The nonlinear system as given in (\ref
{eq:f5})-(\ref{eq:f10}) consists of six independent 
equations in the eight variables $\rho, p_r, p_t, \Delta, E,\sigma$, \textit{y} and 
\textit{Z}. To solve (\ref{eq:f5})-(\ref{eq:f10}) we need to specify two of the quantities involved in the 
integration process.

\section{Choice of potentials \label{choice}}
We need to solve the Einstein-Maxwell field equations (\ref{eq:f5})-(\ref{eq:f10}) by choosing specific forms for the gravitational
 potential $Z$ and the electric field intensity $E$ which are physically reasonable. Then 
 equation (\ref{eq:f9}) becomes a first order equation in the potential $y$ which is integrable.
 
We make the choice
\begin{eqnarray}
Z = \frac{1+(a-b)x}{1+ax} \label{S1},
\end{eqnarray}
where $a$ and $b$ are real constants. The quantity $Z$ is regular at the stellar centre and continuous in the interior 
because of freedom provided  by parameters $a$ and $b$. It is important to realise that this choice for $Z$ is physically
 reasonable and contains special cases which contain neutron star models. When $a = b = 1$ we regain the form of $Z$ 
 for the charged \cite{6} charged stars. For the value $a=1$ the potential corresponds to the \cite{10}
  compact spheres in electric fields. The choice (\ref{S1}) was made by Finch and Skea (1989) to 
 generate stellar models that satisfy all physical criteria for a stellar source. If we set $a=1$, $b=-3/2$ then
  we generate the \cite{20} neutron star model. When $a=7$, $b=8$ then
   we generate the gravitational potential of the \cite{24} superdense stars. Thus the form $Z$ chosen 
   is likely to produce physically reasonable models for charged anisotropic spheres with an equation of state.
   
For the electric field we make the choice
\begin{eqnarray}
\frac{E^{2}}{C} = \frac{k(3+ax)+sa^{2}x^{2}}{(1+ax)^{2}} \label{S2},
\end{eqnarray}
which has desirable physical features in the stellar interior. It is finite at the centre of the star and remains 
bounded and continuous in the interior; for large values of $x$ it approaches a constant value. When $s=0$ then 
we regain $E$ studied by  \cite{8}. However their choice is not suitable as the proper
 charge density becomes  singular at the origin as pointed out by  \cite{18}. Consequently
  we have adapted the form of $E$ so that the proper charge density remains regular throughout the stellar 
  interior with an equation of state. These features become clear in the analysis that follows.

\section{New models\label{models}}
On substituting 
(\ref{S1}) and (\ref{S2}) into (\ref{eq:f9}) we get the first 
order equation
\begin{eqnarray}
\frac{\dot{y}}{y} = \frac{(1+\alpha)b}{4[1+(a-b)x]}+\frac{{\alpha}b}{2(1+ax)[1+(a-b)x]} \nonumber \\
-\frac{\beta(1+ax)}{4C[1+(a-b)x]}\nonumber\\
-\frac{(1+\alpha)[k(3+ax)+sa^{2}x^{2}]}{8(1+ax)[1+(a-b)x]}.\label{S3}
\end{eqnarray}
For the integration of equation (\ref{S3}) it is convenient to consider three cases: $b=0, a=b$ and $a
\neq b$.
\subsection{The case $b=0$\label{model1}}
When $b=0$, (\ref{S3}) gives the solution
\begin{eqnarray}
y=D\left(1+ax\right)^{(-(k-2s)(1+\alpha))/(8a)} \nonumber \\
\times \exp\left[\frac{(1+\alpha)[2k-sax(2+ax)]}{8a(1+ax)}-\frac
{\beta x}{4C}\right], \label{C2}
\end{eqnarray}
where \textit{D} is the constant of integration. The potential \textit{y} in (\ref{C2}) generates a negative 
density $\rho=-\frac{E^{2}}{2}$ which is physically 
undesirable. 
\subsection{The case $b=a$\label{model2}}
When $a=b$, (\ref{S3}) yields the solution
\begin{equation}
y=D\left(1+ax\right)^{(4a\alpha-(2k+s)(1+\alpha))/(8a)}\exp\left[F(x)\right],\label{X2}
\end{equation}
where we have let
\begin{eqnarray}
 F(x) &=& \frac{x}{16C}\left[2C(s-k)(1+\alpha)-4\beta\right.\nonumber\\
& &\left. -a(C(-4+sx)(1+\alpha)+2\beta x)\right], 
\end{eqnarray}
and \textit{D} is the constant of integration. Then we can generate an exact model for the system (\ref
{eq:f5})-(\ref{eq:f10}) in the form
\begin{eqnarray}
\label{X3a}
 e^{2\lambda} = 1+ax, & & \\
\label{X3b}
 e^{2\nu} = A^{2}D^{2}\left(1+ax \right)^{(4a\alpha-(s+2k)(1+ \alpha))/(4a)}\exp\left[2F(x)\right], & &\\
\label{X3c}
 \frac{\rho}{C} = \frac{(2a-k)(3+ax)-sa^{2}x^{2}}{2(1+ax)^{2}}, \qquad \rho > 0,\\
\label{X3d}
 p_{r} = \alpha\rho-\beta, \\
\label{X3e}
 p_{t} = p_{r}+\Delta, \\
\label{X3f}
 \Delta =  & & \nonumber\\
\frac{1}{16C(1+ax)^{3}}\{C^{2}[k^{2}((1+ \alpha)^{2}x(3+ax)^{2}+2sx(1+ax)) 
\nonumber\\
  +4a^{2}x(3-8\alpha+9\alpha^{2}+ a^{2}(1+\alpha)^{2}x^{2}+2ax(2+3\alpha+3\alpha^{2}))\nonumber\\
 -4k(12+a^{3}(1+\alpha)^{2}x^{3}+a^{2}x^{2}(7+9 \alpha +6 \alpha^{2})\nonumber\\
 +ax(12+5\alpha+9\alpha^{2}+4s))\nonumber\\
 +s^{2}(2x(a^{2}x^{2}-1)(2k-asx+s)+a^{2}x^{2}-6ax+1)\nonumber\\
 -2s(ax(6\alpha^{2}+6\alpha+2a(2\alpha^{2}+2\alpha+1)+k(7-2\alpha-\alpha^{2}))\nonumber\\
 +a^{2}x^{2}(\alpha^{2}+2\alpha-2a(1+\alpha+x)+17)+a(\alpha^{2}-2\alpha)\nonumber\\
 -k(1+\alpha)^{2}+7)]-4Cx(1+ax)^{2}[(1+ \alpha)(2a^{2}x-3k-2s)\nonumber\\
 -a\beta(k(1+\alpha)+sax(1+\alpha)-6\alpha-4)]+4\beta^{2}x(1+ax)^{4}\},\\
\label{X3g}
 \frac{E^{2}}{C}= \frac{k(3+ax)+sa^{2}x^{2}}{(1+ax)^{2}},\\
\label{X3h}
 \frac{\sigma^{2}}{C} = \frac{C\left(\sqrt{k}(a^{2}x^{2}+3ax+6)+2\sqrt{s}ax\sqrt{3+ax}(2+ax)\right)^{2}}
{x(3+ax)(1+ax)^{5}}.
\end{eqnarray}
The new exact solution (\ref{X3a})-(\ref{X3h}) of the Einstein-Maxwell system is presented in terms of 
elementary functions. When $s=0$ we regain the first 
class of charged anisotropic models of \cite{8}. For our models the mass function is given by
\begin{eqnarray}
 M(x)=\frac{1}{8C^{3/2}}\left[\frac{(4a^{2}-2ak)x^{3/2}}{a(1+ax)} \right. \nonumber\\
 \left.  + \frac{ s(15+10ax-2a^{2}x^{2})x^{1/2}}{3a(1+ax)}\right.\nonumber\\
 \left.-\frac{5s\arctan(\sqrt{ax})}{a^{3/2}}\right].\label{Gmass1}
\end{eqnarray}
The gravitational potentials and matter variables are well behaved in the interior of sphere. However, as 
in earlier treatments, the singularity in the charge 
distribution at the centre is still present in general. In our new solution the 
singularity can be eliminated when $k=0$. Then equation (\ref{X3h}) becomes
\begin{equation}
\frac{\sigma^{2}}{C} = \frac{4Csa^{2}x(2+ax)^{2}}{(1+ax)^{5}}.\label{ACACIA1}
\end{equation}
At the stellar centre $x=0$ and the charge density vanishes.

\subsection{The case $b\neq a$\label{model3}}
On integrating (\ref{S3}), with $b\neq a$ we obtain
\begin{eqnarray}
y &=& D(1+ax)^{m}[1+(a-b)x]^{n}\nonumber\\
& & \times \exp\left[ -\frac{ax[Cs(1+\alpha)+2\beta]}{8C(a-b)}\right],\label{S9}
\end{eqnarray}
where \textit{D} is the constant of integration. The constants \textit{m} and \textit{n} are given by
\begin{eqnarray}
 m &=& \frac{4\alpha b-(1+\alpha)(s+2k)}{8b},\nonumber\\
 n &=& \frac{1}{8bC(a-b)^{2}}[a^{2}C((1+\alpha)(s+2k)-4\alpha b)\nonumber\\
& & -abC(5k(1+\alpha) \nonumber \\
& & -2b(1+5\alpha))
 +b^{2}(3kC(1+\alpha)\nonumber\\
 & & -2bC(1+3\alpha)+2\beta)].\nonumber
\end{eqnarray}
Then we can generate an exact model for the system (\ref{eq:f5})-(\ref{eq:f10}) in the form\\

\begin{eqnarray}
\label{S10a}
 e^{2\lambda} = \frac{1+ax}{1+(a-b)x},\\
\label{S10b}
 e^{2\nu} = A^{2}D^{2}\left(1+ax \right)^{2m}[1+(a-b)x]^{2n}\nonumber\\
\times\exp\left[-\frac{ax[ Cs(1+\alpha)+2\beta ]}
{4C(a-b)}\right],\\
\label{S10c}
 \frac{\rho}{C} = \frac{(2b-k)(3+ax)-sa^{2}x^{2}}{2(1+ax)^{2}}, \qquad \rho > 0,\\
\label{S10d}
 p_{r} = \alpha\rho-\beta, \\
\label{S10e}
 p_{t} = p_{r}+\Delta, \\
\label{S10f}
 \Delta = \frac{-bC}{(1+ax)}-\frac{bC(1+5\alpha)}{(1+\alpha)(1+ax)^{2}}\nonumber\\
+\frac{2\beta}{1+\alpha}+\frac
{Cx[1+(a-b)x]}{(1+ax)}\nonumber\\
\times
\{4\left(\frac{a^{2}m(m-1)}{(1+ax)^{2}}+\frac{2a(a-b)mn}{(1+ax)[1+(a-b)x]}\right)\nonumber\\
+4\left(\frac{(a-b)^{2}n(n-1)}{[1+
(a-b)x]^{2}}\right)\nonumber\\
-\frac{a[Cs(1+\alpha)+2\beta](a(m+n)[1+(a-b)x]-bn)}{(a-b)C(1+ax)[1+(a-b)x]}\nonumber\\
 +\frac{a^{2}[Cs(1+\alpha)+2\beta]^{2}}{16C^{2}(a-b)^{2}}\}\nonumber\\
-\frac{4[1+ax(2+(a-b)x)]-b(5+
\alpha)x}{4(a-b)(1+\alpha)(1+ax)^{3}[1+(a-b)x]}\nonumber\\
 \times
[-8b^{2}Cn+a^{3}x(-8C(m+n)+[Cs(1+\alpha)+2\beta]x)\nonumber\\
 +a^{2}(8C(m+n)(2bx-1) +[Cs(1+\alpha)+2\beta](2-bx)x)\nonumber\\
 +a(-8b^{2}C(m+n)x+[Cs(1+\alpha) +2\beta]\nonumber\\
 +b(8Cm+16Cn-[Cs(1+\alpha)+2\beta]x))],\\
\label{S10g}
 \frac{E^{2}}{C}= \frac{k(3+ax)+sa^{2}x^{2}}{(1+ax)^{2}},\\
\label{S10h}
\frac{\sigma^{2}}{C} = \frac{C[1+(a-b)x]}{x(3+ax)(1+ax)^{5}}\nonumber\\
\times \left(\sqrt{k}(a^{2}x^{2}+3ax+6) \right. \nonumber\\
 \left. +2\sqrt{s}ax\sqrt{3+ax}(2+ax)
\right).
\end{eqnarray}
The exact solution (\ref{S10a})-(\ref{S10h}) of the Einstein-Maxwell system is written in terms of 
elementary functions.
For this case the mass function is given by
\begin{eqnarray}
 M(x)=\frac{1}{8C^{3/2}}\left[\frac{(4ab-2ak)x^{3/2}}{a(1+ax)} \right. \nonumber\\
 \left. +  \frac{ s(15+10ax-2a^{2}x^{2})x^{1/2}}{3a(1+ax)}\right.\nonumber\\
 \left.-\frac{5s\arctan(\sqrt{ax})}{a^{3/2}}\right].\label{S11}
\end{eqnarray}
This solution is a generalisation of the second class of charged anisotropic models of 
\cite{8}: When $s=0$ we obtain their 
expressions for the gravitational potentials and matter variables. These 
quantities are well behaved and regular in the interior of the sphere. However in general there is a 
singularity in the charge density at the centre. This 
singularity is eliminated when $k=0$ so that
\begin{equation}
\frac{\sigma^{2}}{C} = \frac{4Csa^{2}x[1+(a-b)x](2+ax)^{2}}{(1+ax)^{5}}.\label{ACACIA4}
\end{equation}
At the centre of the star $x=0$ and the charge density vanishes.

\section{Physical features\label{features}}
\begin{figure}
\includegraphics[angle = 0,scale = 0.40]{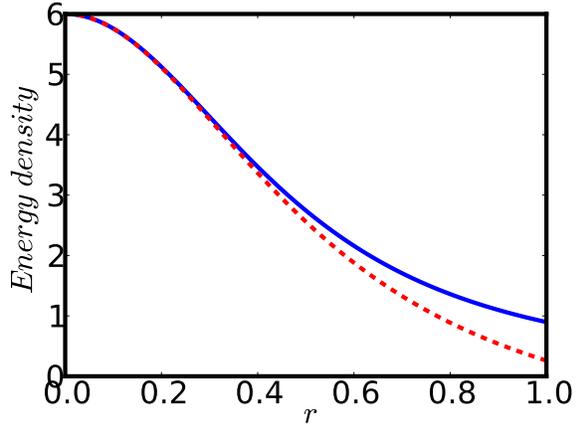}
\caption{Energy density ($\rho$) versus radius. }
\label{fig1}
\end{figure}

\begin{figure}
\includegraphics[angle = 0,scale = 0.40]{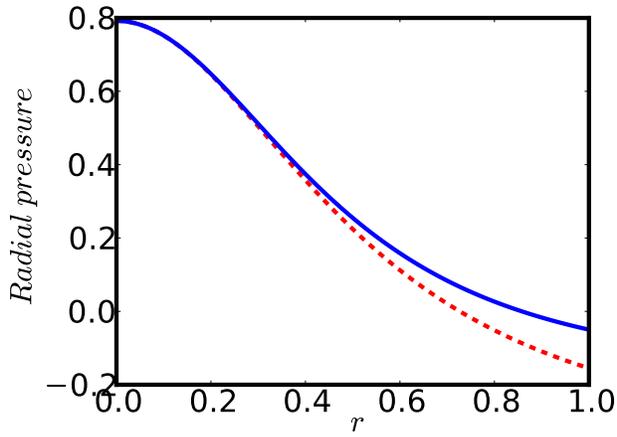}
\caption{Radial pressure ($p_{r}$) versus radius.  }
\label{fig2}
\end{figure}

\begin{figure}
\includegraphics[angle = 0,scale = 0.42]{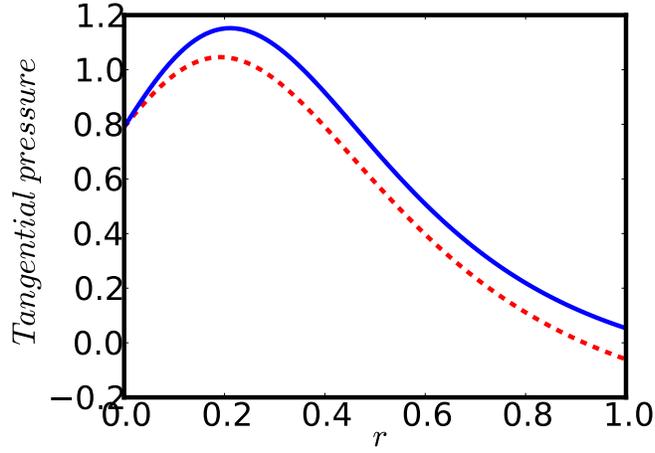}
\caption{ Tangential pressure ($p_{t}$) versus radius.  }
\label{fig3}
\end{figure}

\begin{figure}
\includegraphics[angle = 0,scale = 0.40]{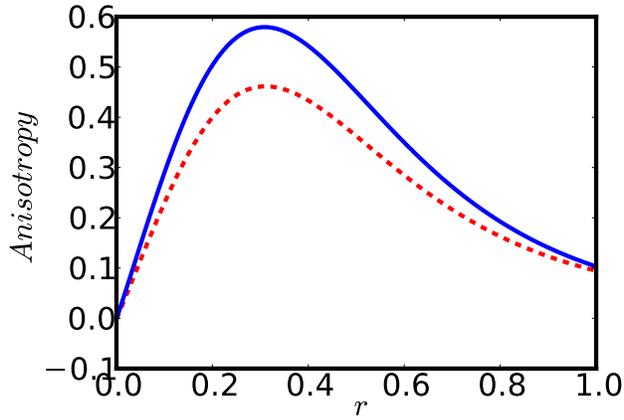}
\caption{Anisotropy ($\Delta$) versus radius.   }
\label{fig4}
\end{figure}

\begin{figure}
\includegraphics[angle = 0,scale = 0.42]{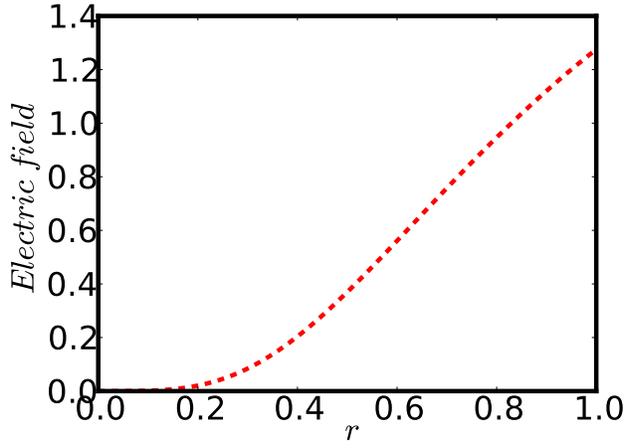}
\caption{Electric field ($E^{2}$) versus radius.}
\label{fig5}
\end{figure}

\begin{figure}
\includegraphics[angle = 0,scale = 0.40]{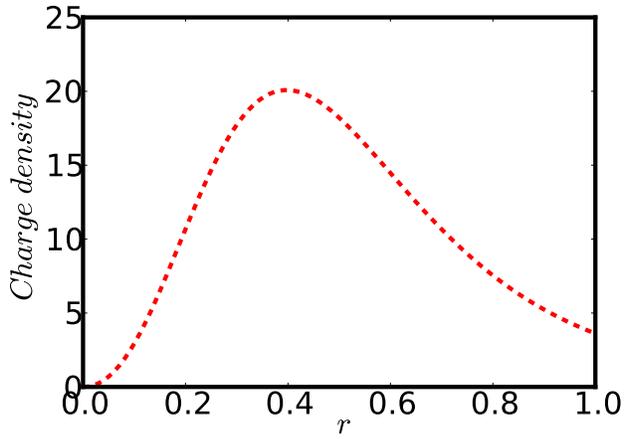}
\caption{Charge density ($\sigma^{2}$) versus radius.}
\label{fig6}
\end{figure}

\begin{figure}
\includegraphics[angle = 0,scale = 0.40]{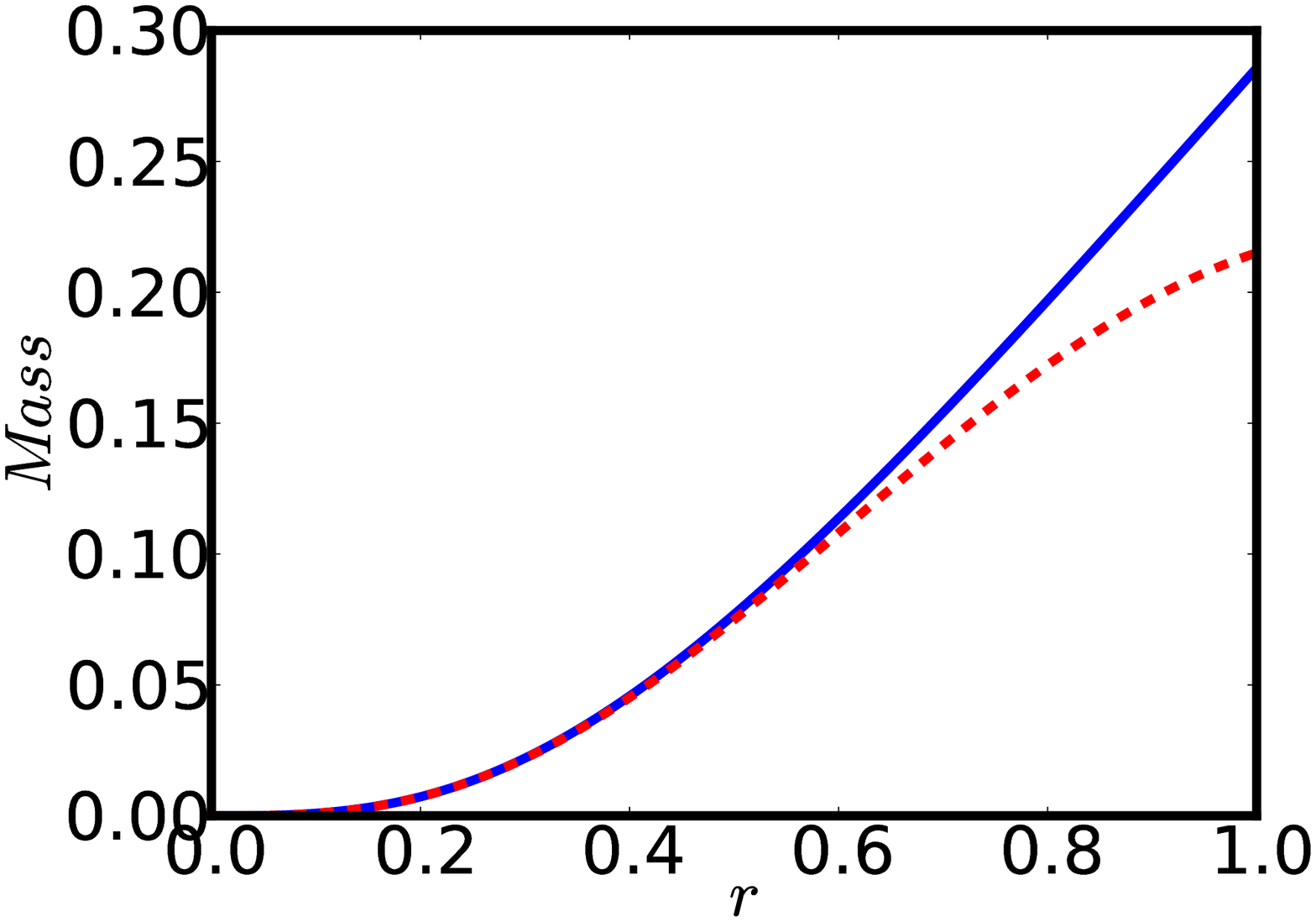}
\caption{Mass ($M$) versus radius.}
\label{fig7}
\end{figure}

In this section we show that the exact solutions found in section \ref{models}, for particular choices of the parameters $a,~b$ and $s$, are 
physically reasonable. A detailed physical analysis for general values of the parameters will be a future investigations.
We used the programming language Python to generate these plots for case: $a>b$ ($a=2.5$, $b=2.0$), $\alpha = 0.33$, $\beta = \alpha
\tilde{\rho}=0.198$, $C=1$ and $s=2.5$, $k=0$ where $\tilde{\rho}$ is 
the density at the boundary. The solid lines correspond to $s=0$ and the dashed lines correspond to $s\neq0$ in the graphs.
 We generated the following plots: energy density (Figure \ref{fig1}), radial pressure (Figure \ref{fig2}), tangential pressure (Figure \ref{fig3}), 
anisotropy measure (Figure \ref{fig4}), electric field intensity (Figure \ref{fig5}), charge density (Figure \ref{fig6}) and mass (Figure \ref{fig7}), 
  The energy density $\rho$ is positive, finite and monotonically 
decreasing. The radial pressure $p_{r}$ is similar to $\rho$ since $p_{r}$ and $\rho$ are related by a 
linear equation of state. The values of $\rho$ and $p_{r}$ are lower in the presence of the electric field $E\neq0$. 
The tangential pressure is well behaved increasing away from the centre, reaches a maximum and becomes a decreasing function. 
This is reasonable since the conservation of angular 
momentum during the quasi-equilibrium contraction of a massive body should lead to high values of $p_{t}$ in central regions of the star as pointed out by \cite{9}. 
The anisotropy  $\Delta$ is increasing in the neighbourhood of the centre, reaches a maximum value and then subsquently decreases. 
The profile of $\Delta$ is similar  to the \cite{27}  and the \cite{125}  for strange stars with quark matter.
 
The form chosen for \textit{E} is physically reasonable and describes a function which is initially small and then increases as we approach the 
boundary. The charge density in general is continuous, initially increases and then 
decreases. Note that the singularity at the stellar centre is eliminated since $k=0$. The mass 
function is strictly increasing function which is continuous and finite. We observe that the mass, in the presence of charge, 
has lower values that the corresponding uncharged case. This is consistent as $E\neq0$ generates 
lower densities which produces a weaker total field since 
the electromagnetic field is repulsive. Thus all matter variables, 
electromagnetic quantities and gravitational potentials are nonsingular and well behaved in a region 
away from the stellar centre. We emphasize that the electromagnetic quantities are all 
well behaved close to the stellar centre since
there are finite values for the charge density. This is different from other treatments with an equation
of state.

The solutions found in this paper can be used to model realistic stellar bodies. We introduce the 
transformations:
$\tilde{a}=a R^2,~
\tilde{b}=bR^2,~\tilde{\beta}=\beta R^2,~\tilde{k}=k R^2,~\tilde{s}=s R^{2}.$  Using these transformations 
the energy density becomes
\begin{equation}
\label{eq:f55}\rho =\frac{(2 \tilde{b}-\tilde{k})(3+\tilde{a}y)-\tilde{s}\tilde{a}^{2}y^{2}}{2
R^2 (1+\tilde{a}y)^2}.
\end{equation}
The mass contained within a radius \textit{r} is given by
\begin{eqnarray}
\label{eq:f56}M &=& \frac{r^{3}(6\tilde{b}
-3\tilde{k}+5\tilde{s})}{12R^{2}(1+\tilde{a}y)}+\frac{\tilde{s}r(15-2\tilde{a}y^{2})}{24\tilde{a}(1+\tilde{a}
y)}\nonumber\\
& & -\frac{5\tilde{s}R\arctan[\sqrt{\tilde{a}y}]}{8\tilde{a}^
{3/2}},
\end{eqnarray}
where we have set $C=1$ and $y=\frac{r^2}{R^2}$. If $\tilde{k}=0$, $\tilde{s}=0$ $(E=0)$ then we have
\begin{eqnarray}
\label{eq:f59a}
\rho &=&\frac{\tilde{b}(3+\tilde{a}y)}{R^2(1+\tilde{a}y)^2},\\
\label{eq:f59b}
M &=&\frac{\tilde{b}r^3}{2R^2(1+\tilde{a}y)}.
\end{eqnarray}
In this case there is no charge and we obtain the expressions \cite{27}.
For the astrophysical importance of our solutions, we try to compare the masses corresponding to the 
models of this paper to those found by \cite{27} and \cite{8}.

We calculate the masses for the various cases with and without charge. We set  $r=7.07$ km, 
$R=43.245$ km, $\tilde{k}=37.403$ and $\tilde{s}=0.137$. We 
tabulate the information in Table \ref{Table1}. Note that when $k=0$, 
$s=0$ we have an uncharged stellar body and we regain masses generated by \cite{27}. When $k\neq0$ and $s=0$ we find masses for 
a charged relativistic star in the \cite{8}
models. We have included those two sets of values for consistency and to demonstrate that our general 
results contain the special cases considered 
previously. When $k=0$ and $s\neq0$ the masses found correspond to new 
charged solutions, with nonsingular charge densities at the origin. When $k\neq0$ and $s\neq0$ then 
we have most general case. In all cases we obtain 
stellar masses which are physically reasonable. We observe that the 
presence of charge generates a lower mass \textit{M} because of the repulsive electromagnetic field 
which corresponds to a weaker field. Observe that our 
masses are consistent with the results of \cite{21,22,23}
with an equation of state for strange matter. When the charge is absent the mass $M=1.434 M_
{\bigodot}$; the presence of charge in the different solutions 
affects this value. \cite{21,22,23} have shown that these 
values are consistent with observations for the X-ray binary pulsar SAX J1808.4-3658. Consequently 
these charged general relativistic models have 
astrophysical significance. A trend in the masses is observable in table \ref{Table1}. 
It is interesting to observe the smallest masses are attained when $k\neq0$, $s=0$ in the 
presence of the electromagnetic field. When  $k\neq0$, 
$s=0$  then the electric field intensity is stronger by (\ref{S2}) which 
negates the attraction of the gravitational field leading to a weaker field.
Note that we have also included the value of anisotropy $\Delta$ in table 1. We note that larger stellar masses correspond to increasing values of anisotropy. 

\begin{deluxetable}{ccrrrrrrrrcrl}
\tablecaption{Central density and mass for different
anisotropic stellar models for neutral and charged bodies.\label{tbl-1}}
\tablewidth{0pt}
\tablehead{\colhead{$\tilde{b}$} & \colhead{$\tilde{a}$} & \colhead{$\alpha$}& \colhead{${\rho}_c$} & \colhead{$M$}& \colhead{${\rho}_c$} & \colhead{$M$} & \colhead{$\Delta$} \\
& & & \colhead{$(\times 10^{15}~\mbox{gcm}^{-3})$}& \colhead{$M_{\odot}$}& \colhead{$(\times 10^{15}~\mbox{gcm}^{-3})$} &\colhead{$M_{\odot}$} &  \\
& & & \colhead{$E=0$}& \colhead{$E=0$}& \colhead{$E\neq0$} &\colhead{$E\neq0$}}
\startdata\colhead{30} &\colhead{23.681} &\colhead{0.401} &\colhead{2.579}  &\colhead{1.175}&\colhead{0.971} &\colhead{0.433} &\colhead{0.039}\\
\colhead{40} &\colhead{36.346} &\colhead{0.400}&\colhead{3.439} & \colhead{1.298} &\colhead{1.831} &\colhead{0.691}  &\colhead{0.054} \\
\colhead{50} &\colhead{48.307}&\colhead{0.424}  &\colhead{4.298} &\colhead{1.396} &\colhead{2.691}  &\colhead{0.874}  &\colhead{0.072} \\
\colhead{54.34} &\colhead{53.340} &\colhead{0.437} &\colhead{4.671}&\colhead{1.433} &\colhead{3.064} &\colhead{0.940}  &\colhead{0.081}\\
\colhead{60}& \colhead{59.788} &\colhead{0.457}&\colhead{5.158} &\colhead{1.477} &\colhead{3.550} &\colhead{1.017} &\colhead{0.094} \\
\colhead{70} &\colhead{70.920}&\colhead{0.495} &\colhead{6.017}  &\colhead{1.546} &\colhead{4.410} & \colhead{1.133} &\colhead{0.119} \\
\colhead{80}&\colhead{81.786}&\colhead{0.537} &\colhead{6.877}  &\colhead{1.606} &\colhead{5.269} & \colhead{1.231} &\colhead{0.146} \\
\colhead{90}&\colhead{92.442}&\colhead{0.581} &\colhead{7.737}  &\colhead{1.659} &\colhead{6.129} & \colhead{1.314} &\colhead{0.177} \\
\colhead{100}&\colhead{102.929}&\colhead{0.627} &\colhead{8.596}  &\colhead{1.705} &\colhead{6.989} & \colhead{1.386} &\colhead{0.207} \\
\colhead{183}&\colhead{186.163}&\colhead{1.083} &\colhead{15.730}  &\colhead{1.959} &\colhead{14.124} & \colhead{1.759} &\colhead{0.593} 
\enddata
\end{deluxetable}
\section{\label{Q} Density Variation}
From (\ref{eq:f55}) we observe that
\begin{equation}
\label{mhj1}
 \rho_{c}=\frac{(2\tilde{b}-\tilde{k})(3+\tilde{a}\frac{c^{2}}{R^{2}})-\tilde{s}\tilde{a}^{2}\frac{c^{4}}{R^{4}}}{2R^{2}(1+\tilde{a}\frac{c^{2}}{R^{2}})^{2}},
\end{equation}
is the density at the stellar surface. The density at the centre of the star is
\begin{equation}
\label{mhj2}
 \rho_{0}=\frac{3(2\tilde{b}-\tilde{k})}{2R^{2}}.
\end{equation}
Then we can generate the ratio
\begin{equation}
\label{mhr3}
\epsilon=\frac{(2\tilde{b}-\tilde{k})(3+\tilde{a}\frac{c^{2}}{R^{2}})-\tilde{s}\tilde{a}^{2}\frac{c^{4}}{R^{4}}}{3(2\tilde{b}-\tilde{k})(1+\tilde{a}\frac{c^{2}}{R^{2}})^{2}},
\end{equation}
called the density constrast. With the help of (\ref{mhr3}) we can find $\frac{c^{2}}{R^{2}}$ in terms of $\epsilon$:
\begin{eqnarray}
\label{mhr4}
 \frac{c^{2}}{R^{2}}&=&\frac{-(2\tilde{b}-\tilde{k})(6\epsilon-1)}{2a(3\epsilon (2\tilde{b}-\tilde{k})+\tilde{s})}\nonumber\\
 & & + \frac{\sqrt{(24\epsilon+1)(2\tilde{b}-\tilde{k})^{2}+12\tilde{s}2\tilde{b}-\tilde{k})(1-\epsilon)}}{2a(3\epsilon (2\tilde{b}-\tilde{k})+\tilde{s})}.
\end{eqnarray}
Then (\ref{eq:f56}) and (\ref{mhr4}) yield the quantity
\begin{eqnarray}
\label{mhr5}
\frac{M}{c}&=&\frac{c^{2}(6\tilde{b}
-3\tilde{k}+5\tilde{s})}{12R^{2}(1+\tilde{a}\frac{c^{2}}{R^{2}})}+\frac{\tilde{s}r(15-2\tilde{a}\frac{c^{4}}{R^{4}})}{24\tilde{a}(1+\tilde{a}\frac{c^{2}}{R^{2}})}\nonumber\\
& & -\frac{5\tilde{s}R\arctan[\sqrt{\tilde{a}\frac{c^{2}}{R^{2}}}]}{8c\tilde{a}^{3/2}},
\end{eqnarray}
which is the compactification factor. consequently our model is characterised by the surface density $\rho_{c}$, the density contrast $\epsilon$ and the compactification
 factor $\frac{M}{c}$ in presence of charge. 
These two parameters produce information of astrophysical significance for specific choice of the parameters. 
The compactification factor classifies stellar objects in various categories depending on the range of of $\frac{M}{c}$: for normal stars $\frac{M}{c} \sim 10^{-5}$,
 for white dwarfs $\frac{M}{c} \sim 10^{-3}$,
 for neutron stars $\frac{M}{c} \sim 10^{-1}$ to $\frac{1}{4}$, for ultra-compact stars $\frac{M}{c} \sim \frac{1}{4}$ to $\frac{1}{2}$, and for black holes $\frac{M}{c} \sim \frac{1}{2}$. 
The parameter $\frac{M}{c}$  in the case of strange stars is in the range of ultra-compact stars with matter densities greater than the nuclear density.

\begin{deluxetable}{ccrrrrrrrrcrl}
\tablecaption{Variation of density.\label{tbl-2}}
\tablewidth{0pt}
\tablehead{\colhead{$\tilde{a}$} & \colhead{$\tilde{b}$} & \colhead{$\epsilon$}& \colhead{$c$} & \colhead{$\frac{M}{c}$} & &\colhead{$\frac{M}{M_{\odot}}$}}
\startdata\colhead{1.6} &\colhead{2} &\colhead{0.3} &\colhead{8.30}  &\colhead{0.073} & &\colhead{0.60}\\
\colhead{(1.6)} &\colhead{(0.6)} &\colhead{(0.3)}&\colhead{(10.18)} & \colhead{(0.100)} & &\colhead{ (1.02)} \\
 & & & & & & \\
\colhead{2.4} &\colhead{ 2.8}&\colhead{0.5}  &\colhead{10.67} &\colhead{ 0.093} & &\colhead{0.99} \\
\colhead{(2.4)} &\colhead{(1.4)} &\colhead{(0.5)} &\colhead{(11.20)}&\colhead{(0.102)} & &\colhead{(1.14)}\\
& & & & & & \\
\colhead{6.0}& \colhead{6.4} &\colhead{0.1}&\colhead{13.59} &\colhead{0.277} & &\colhead{3.76}  \\
\colhead{(6.0)} &\colhead{(5.0)}&\colhead{(0.1)} &\colhead{(15.83)}  &\colhead{(0.329)} & &\colhead{(5.20)} 
\enddata
\end{deluxetable}
For particular choices of the parameters $\tilde{a}$, $\tilde{b}$, $\tilde{k}$, $\tilde{s}$ and specifying the density contrast $\epsilon$ we can find the boundary $c$ from (\ref{mhr4}). 
Note that the parameter $R$ is specified if we take the central density to be $\rho_{0}=2\times 10^{14} gm~cm^{-3}$ in (\ref{mhj2}). 
The compactification factor $\frac{M}{c}$ then follows from (\ref{mhr5}).
For charged matter we take the values $k=2.8$ and $s=2$ and for uncharged matter $k=0$ and $s=0$. Table 2 represents typical values for $\epsilon$ and $\frac{M}{c}$
 for charged matter and uncharged matter; the first set of values are for charged matter and the bracketted values are the corresponding values for neutral matter when $\tilde{k}= \tilde{s}=0$.
For uncharged matter we regain the values of $\epsilon$ and $\frac{M}{c}$ generated by \cite{25} for superdense stars models with neutral matter.
We observe that the presence of charge  has the effect of reducing both $c$ and the compactification $\frac{M}{c}$ to produce the same value of the density constrast $\epsilon$.
This is consistent since the presence of the electric field intensity has the effet of leading to a weaker field. 
The values that we have generated in Table 2 for neutral and charged matter permit configurations typical of neutron stars and strange stars. 
Note the changing the various parameters will allow for smaller or larger compactification factors.
Thus the class models found in this paper allow for stellar configurations which provide physically viable models of superdense structures.

\section{Conclusion\label{concl}}
The recent results contained in the investigations of \cite{g1,g2,g3},
\cite{m1,m2,m3} and \cite{k1} have highlighted the importance of including the electric field in gravitatinal compact objects.
This serves as a motivation for our study with equation of state.
The models of \cite{8}, which contain several earlier solutions are regained 
from our solutions in this paper for particular values of 
the parameters \textit{k} and \textit{s}. Our aim in this paper was to 
find new exact solutions to the Einstein-Maxwell systems with a barotropic equation of state for static 
spherically symmetric gravitational fields. In particular we 
chose a linear equation of state relating the energy density to the 
radial pressure. Such models may be used to model relativistic stars in astrophysical situations. The 
charged relativistic solutions to the Einstein-Maxwell 
systems presented are physically reasonable. A graphical analysis has 
shown that the matter and electromagnetic variables are physically reasonable. In particular our 
solutions contain models, corresponding to $k=0$, for which 
the proper charge density is nonsingular. This is an improvement on 
earlier models which possessed a singularity in $\sigma$ at the stellar origin. Our models yield stellar 
masses and densities consistent with \cite{8} in the presence of charge, and \cite{21,22,23}
and \cite{27} in the limit of vanishing charge. Our solutions may be helpful in the 
study of stellar objects such as SAX J1804.4-3658. We 
have regained the corresponding mass $M= 1.434 M_{\odot}$ 
when $k=0$, $s=0$ of earlier treatments. These models may be useful in the description of charged 
anisotropic bodies, quark stars and configurations with 
strange matter.

\acknowledgments
PMT thanks the National Research Foundation and the University of
KwaZulu-Natal for financial support. SDM acknowledges that this
work is based upon research supported by the South African Research
Chair Initiative of the Department of Science and
Technology and the National Research Foundation.

%\nocite{*}
 %\bibliographystyle{spr-mp-nameyear-cnd}
% %\bibliography{myref}
% \bibliography{biblio-u1}

\end{document}